\documentclass[a4paper,twoside]{article}

\usepackage{graphicx} % Required for inserting images
\usepackage{hyperref}
\usepackage{nomencl}
\usepackage{blindtext}
\usepackage{xcolor}
\usepackage{multirow}
\usepackage{epsfig}
\usepackage{subcaption}
\usepackage{calc}
\usepackage{amssymb}
\usepackage{amstext}
\usepackage{amsmath}
\usepackage{amsthm}
\usepackage{multicol}
\usepackage{pslatex}
\usepackage{apalike}
\usepackage{algorithm2e}
\usepackage{braket}
\usepackage[bottom]{footmisc}
\usepackage{SCITEPRESS}     % Please add other packages that you may need BEFORE the SCITEPRESS.sty package.

\newcommand{\ctpkwh}{\text{ct}/\text{kWh}}
\newcommand{\ctpkwhs}{\text{ct}/\text{kWh}^2}

\begin{document}

\title{Grid Cost Allocation in Peer-to-Peer Electricity Markets: Benchmarking Classical and Quantum Optimization Approaches}

\author{%
\authorname{David Bucher\sup{1}\orcidAuthor{0009-0002-0764-9606},
Daniel Porawski\sup{1},
Benedikt Wimmer\sup{1},
Jonas Nüßlein\sup{2}\orcidAuthor{0000-0001-7129-1237},
Corey O'Meara\sup{3}\orcidAuthor{0000-0001-7056-7545},
Giorgio Cortiana\sup{3}\orcidAuthor{0000-0001-8745-5021},
and Claudia Linnhoff-Popien\sup{2}
}
\affiliation{\sup{1}Aqarios GmbH, Prinzregentenstraße 120, 81677 Munich, Germany}
\affiliation{\sup{2}Department for Computer Science, LMU Munich, Germany}
\affiliation{\sup{3}E.ON Digital Technology GmbH, Hannover, Germany}
\email{\{david.bucher, daniel.porawski\}@aqarios.com,\\\{jonas.nuesslein, linnhoff\}@ifi.lmu.de,\\\{corey.o'meara, giorgio.cortiana\}@eon.com}
}


\abstract{
This paper presents a novel optimization approach for allocating grid operation costs in Peer-to-Peer (P2P) electricity markets using Quantum Computing (QC). We develop a Quadratic Unconstrained Binary Optimization (QUBO) model that matches logical power flows between producer-consumer pairs with the physical power flow to distribute grid usage costs fairly. The model is evaluated on IEEE test cases with up to 57 nodes, comparing Quantum Annealing (QA), hybrid quantum-classical algorithms, and classical optimization approaches. Our results show that while the model effectively allocates grid operation costs, QA performs poorly in comparison despite extensive hyperparameter optimization. The classical branch-and-cut method outperforms all solvers, including classical heuristics, and shows the most advantageous scaling behavior. The findings may suggest that binary least-squares optimization problems may not be suitable candidates for near-term quantum utility.}

\onecolumn \maketitle \normalsize \setcounter{footnote}{0} \vfill

%%%%%%%%%%%%%%%%%%%%%%%%

% \setlength{\nomitemsep}{-\parsep}
% \setlength{\nomlabelwidth}{1.5cm}
% \nomenclature[10]{\(G(V, E)\)}{Power grid graph ($V$ busses, $E$ lines).}
% \nomenclature[20]{\(P\)}{Producer set $P \subset V$}
% \nomenclature[30]{\(C\)}{Consumer set $C \subset V$}
% \nomenclature[40]{\(w_{i,j}\)}{Directed power flow over line $(i,j) \in E$}
% \nomenclature[50]{\(d^i\)}{Production/Consumption volume of node $i$. For producer $d^p < 0, p \in P$. For consumer $d^c > 0, c \in C$}.
% \nomenclature[60]{\(d^{p,c}\)}{Trade volume for pair $d^{p,c} = \min\{-d^p, d^c\}$}  
% \nomenclature[70]{\(u_{i,j}\)}{Utilization (cost factor) of line $(i, j)$}
% \nomenclature[80]{\(v_{i,j}^{p,c}\)}{Directed power flow over line $(i, j) \in E$ if only considering $(p,c)$.}
% \nomenclature[90]{\(v_{i,j}(x)\)}{Sum of all active pair power flows between node $i$ and $j$.}
% \nomenclature[91]{\(M^{p,c}\)}{Cost allocation for trade $(p, c)$}
% \nomenclature[92]{\(\alpha\)}{Power flow matching penalty $[\ctpkwhs]$}
% \nomenclature[93]{\(\rho\)}{Grid fee parameter $[\ctpkwh]$}
% \printnomenclature

\section{Introduction}

The increasing adoption of distributed energy resources and the ongoing transformation of electricity consumers into prosumers drive fundamental changes in power systems operation. P2P electricity markets have emerged as a promising paradigm to facilitate direct energy trading between prosumers while maintaining grid stability and operational efficiency~\cite{sousa2019}. However, a key challenge in implementing P2P markets lies in fairly allocating grid operation costs among participants based on their actual infrastructure usage. Traditional constant network tariffs become inadequate in P2P settings as they fail to account for the complex power flow patterns that emerge from bilateral trades~\cite{baroche2019}, leading to unfair cost distributions between customers. This requires dynamic cost allocation mechanisms to appropriately distribute grid operation and maintenance costs among market participants based on their contribution to network utilization. To implement P2P, the Distribution System Operator (DSO) requires that the cost allocation mechanism covers all operational and maintenance costs. The combinatorial nature of matching multiple producers and consumers presents a computationally challenging optimization problem.

QC, particularly Quantum Annealing (QA), has shown promise in solving complex combinatorial optimization problems in various domains~\cite{abbas2023}. Recent advances in quantum hardware and hybrid quantum-classical algorithms offer new opportunities to address challenging energy sector optimization problems~\cite{blenninger2024}. However, the practical utility of quantum approaches for P2P market optimization remains largely unexplored.

This paper makes two main contributions: First, we develop a novel optimization model for allocating grid usage costs in P2P electricity markets based on actual power flow patterns. Second, we conduct a comprehensive benchmark study comparing quantum annealing, hybrid quantum-classical, and classical optimization approaches for solving the proposed model. Our analysis provides insights into the potential and limitations of current quantum optimization techniques for P2P market applications.

The remainder of this paper is organized as follows: Sec.~\ref{sec:background} provides background on quantum annealing and classical optimization and reviews related work. Sec.~\ref{sec:formulation} presents our problem formulation for P2P cost allocation. Sec.~\ref{sec:case_study} analyzes the model behavior through a detailed case study. Sec.~\ref{sec:benchmarks} presents benchmark results comparing different optimization approaches, and Sec.~\ref{sec:conclusion} concludes, giving future research directions.

% Contents
% \begin{itemize}
%     \item Introduction P2P Market Model: Future candidate for electricity markets
%     \item Grid Operator requires operational costs to be covered: (1) line operation costs, and (2) grid stabilization costs.
%     \item Dynamic way of billing customers how \emph{strong} they are using the grid
% \end{itemize}
\section{Background \& Related Work}\label{sec:background}
% paragraph tansition
\subsection{Quantum Annealing}

The quantum bit (qubit), as the elementary information unit in quantum mechanics, manifests in superposition states $\ket{\psi} = \alpha \ket{0} + \beta \ket{1}$ with $|\alpha|^2 + |\beta|^2 = 1$. This property enables $n$ coupled qubits to encode the complete set of $2^n$ bit configurations in a quantum state described by
\begin{align}
\ket{\psi} = \sum_{x = 0}^{2^n - 1} \psi_x \ket{x},
\end{align}
for $x \in \{0, 1\}^n$. Computational basis measurements reveal bit string $x$ with probability $|\psi_x|^2$~\cite{nielsen2010}, forming the foundation for quantum optimization approaches that maximize the measurement probability of optimal bit configurations.

The framework for quantum optimization typically employs an Ising Hamiltonian constructed from Pauli $z$-matrices $\hat \sigma^z_i = \ket{0}\bra{0} - \ket{1}\bra{1}$. This cost operator takes the form
\begin{align}\label{eq:ising_model}
\hat H_{C} = -\sum_{i, j} J_{i,j} \hat \sigma^z_i \otimes \hat \sigma^z_j -\sum_{i} h_i \hat \sigma^z_i,
\end{align}
where tensor products indicate simultaneous operations on qubit pairs $(i,j)$. The diagonality of $\hat H_{C}$ in the $z$-basis ensures its ground state corresponds to a computational basis state $x^*$.
The classical representation of bit string energies follows naturally as $\bra{x}\hat H_C \ket{x} = H_C(x)$, yielding the QUBO formulation:
\begin{align}
H_C(x) = -\sum_{i,j}J_{i,j} z_i z_j  - \sum_i h_i z_i = \sum_{i,j} Q_{i,j} x_i x_j,
\end{align}
with $z_i = 1 - 2 x_i$ and $\hat \sigma^z_i \ket{x_i} = z_i \ket{x_i}$. This equivalence enables physical implementation of the Ising Hamiltonian for combinatorial optimization~\cite{lucas2014b}.
The solution methodology leverages the adiabatic theorem~\cite{born1928,farhi2001}, which postulates that a system maintains ground-state occupation under sufficiently slow Hamiltonian evolution. Adiabatic Quantum Computing (AQC) exploits this principle by initializing the system in $\ket{+} = 2^{-n/2}\sum_x \ket{x}$, the ground state of $\hat H_D = - \sum_i \hat \sigma^x_i$. Time evolution proceeds under
\begin{align}\label{eq:time-evolution}
\hat H(s) = s \hat H_C + (1 - s) \hat H_D,
\end{align}
with $s \in [0, 1]$. Ideally, adiabatic evolution from $s = 0$ to $s = 1$ yields the QUBO solution state $\ket{x^*}$.

Physical implementations, however, must contend with decoherence and noise, necessitating accelerated evolution times. This practical constraint transitions the process from AQC to QA~\cite{rajak2023}. While this departure from the adiabatic regime reduces the optimal state probability $|\psi_{x^*}|^2$, repeated measurements can still reveal the desired solution. Hence, QA is considered approximate optimization.

Current QA hardware, exemplified by D-Wave's Advantage System QPU, implements this approach using over 5000 qubits interconnected by 35000 couplers~\cite{mcgeochb}. The underlying Pegasus topology establishes 15 couplings per qubit, though this fixed architecture introduces implementation challenges for problems requiring higher connectivity. Such cases necessitate the embedding of logical qubits as multiple physical qubits (so-called chains).

In addition to its QPUs, D-Wave offers a cloud-based hybrid quantum computing platform (Leap) that utilizes both quantum and classical computing to accelerate the solution process using a proprietary algorithm.

\subsection{Classical Solvers}

Optimization heuristics produce fast solutions repetitively to arrive at a reasonable solution candidate instead of a structured search in the space of possible solutions. Even though the optimal solution is often not reachable for heuristics, the fast runtime and close-to-optimal solutions are more desirable in many practical application scenarios, especially if exact methods suffer from lengthy runtimes.

Simulated Annealing (SA) is a QUBO heuristic, a Markov-Chain Monte Carlo method, that starts from a random initial point and iteratively proposes bit-flips in its current solution~\cite{kirkpatrick1983a}. If a bit-flip decreases cost, it will be accepted directly. Still, when it increases, it will only be accepted based on an acceptance probability dictated by the cost delta and an annealing parameter $\beta$. This parameter increases with every iteration, shifting the search strategy from exploration to exploitation.

Tabu Search (TS) enhances the local search method by maintaining a \emph{tabu list} of recently explored samples in memory to prevent cycling and escape local optima~\cite{glover1989}. At each iteration, it evaluates the neighborhood of the current solution and moves to the best non-tabu neighbor, even if this leads to a temporary deterioration in solution quality.

Gurobi is a state-of-the-art classical branch-and-cut solver that structurally explores the search space using the branch-and-bound algorithm in conjunction with various cutting plane methods~\cite{gurobi}. Gurobi is capable of solving linear programming as well as quadratic programming problems with various constraints.

\subsection{Related Work}

To the best of the authors knowledge, QC has not been investigated to solve the problem of P2P Cost Allocation directly, although several applications of QC within the optimal use of energy exist. Specifically, applications of QA to solve the problem of grid partitioning use-cases~\cite{fernandez-campoamor2021c,colucci2023,bucher2024} have successfully demonstrated the some of these first applications. Furthermore, gate-based optimization algorithms showed promising results for applications in load scheduling~\cite{mastroianni2024a}, charging optimization~\cite{kea2023a}, and coalition formation~\cite{mohseni2024}. Finally, QA has been applied for matching producer and consumer pairs in P2P markets~\cite{omeara2023} without considering cost allocation.

P2P markets have been studied extensively in the literature. Early work on P2P electricity markets focused primarily on bilateral trading mechanisms where prosumers directly negotiate energy exchanges. Later work introduced a P2P energy trading framework incorporating network constraints and distribution fees~\cite{sousa2019}. The challenge of fair cost allocation gained prominence as researchers recognized that traditional volumetric network tariffs could lead to inefficient market outcomes in P2P settings.
This study is based on modeling aspects developed by Baroche et al.~\cite{baroche2019}, who applied cooperative game theory concepts like the Shapley value to allocate network costs among peers, and has been extended by incorporating network losses and congestion costs in their allocation framework~\cite{lecadre2020}.
%Morstyn et al.~\cite{morstyn2018} proposed a bilateral contract networks approach that explicitly considered the physical power flows resulting from P2P trades.

\section{Problem Formulation}\label{sec:formulation}
A power grid can be described as a graph $G(V, E)$ with power lines as edges $E$ and loads or generators as nodes $i \in V = P \cup C$, where $P$ and $C$ are the sets of producers and consumers, respectively.  
Given a self-reliant section of the power grid with the nodal power injection $d^i$ balanced by the load, i.e., $\sum_i d^i = 0$, we strive to find an assignment $z = \{(p_1, c_1), (p_2, c_2), \dots\} \subseteq P \times C$, such that the following condition is satisfied: The combination of all pair power flows between the pairs in $z$---called \emph{logical power flow} in the following---should match the physical power flow, which is the power flow when all participants inject and draw their full load. By doing so, we can approximately locate which participant is responsible for which load in the network power lines. In reality, many possible assignments can lead to a close matching of the power flows. Since line usage is attributed to cost, we want to identify the assignment that benefits the customers most, i.e., minimizes overall costs for the customers.

Therefore, we devise a multicriteria optimization problem from the combination of two objectives: The first objective is to minimize the mismatch between physical and logical power flow, while the second objective is to minimize attributed costs.

\subsection{Power Flow Matching}
Given the power generators and loads $d^i$ and the grid infrastructure $G(E, V)$, we can compute the baseline power flow $w_{i,j}\,\forall\, (i, j) \in E$ through standard methods like DC power flow~\cite{thurner2018a}.
The power flow is directed in the sense that $w_{i,j} = -w_{j,i}$. Furthermore, we consider the optimization within a fixed time period $\Delta t$. Thus, power flow and injection quantities are referred to in energy units, not in power units ($[\text{kWh}]$ instead of $[\text{kW}]$).

The pair power flow $v_{i,j}^{p,c}$ can analogously be computed using DC power flow, but instead of all nodes being active and injecting (drawing) power into (from) the grid, we only enable the nodes $p$ and $c$. Furthermore, we match the production capacity of $p$ to the load demand of $c$, by using $d^{p,c} = \min\{-d^p, d^c\}$ for both nodes.
This computation has to be repeated for all possible pairs $P \times C$ to obtain the possible power pair flows. The definition is comparable to the electrical power transfer distance, defined in Refs.~\cite{baroche2019,christie2000}.

Using the binary variables $x_{p,c} \in \{0, 1\}$ to indicate whether a pair is part of the final assignment $z$, we can describe the \emph{logical power flow} as the linear superposition of all active pair power flows
\begin{align}\label{eq:logical-pf}
   v_{i,j}(x) = \sum_{p,c} v_{i,j}^{p,c}x_{p,c}.
\end{align}

As a consequence, we can model the objective to minimize the mismatch between logical and baseline power flow as a convex least-squares optimization problem
\begin{align}\label{eq:pf-matching}
    \min_x \sum_{(i,j) \in E} \left(w_{i,j} - v_{i,j}(x)\right)^2.
\end{align}
The solution $x^*$ indicates which producer-consumer pairs have been selected for P2P trading.

\subsection{Cost allocation}

We can estimate the line utilization through $u_{i,j} = |w_{i,j}|/W_{i,j}$, where $W_{i,j}$ is the specific maximum capacity constant of a power line. To charge the usage of a line, we employ the utilization as a cost factor and bill the power flow over a line in accordance to a grid fee constant $\rho$ [$\text{ct}$/$\text{kWh}$]:
\begin{align}\label{eq:cost-allocation-1}
    M^{p,c} = \rho \sum_{(i,j) \in E} u_{i,j} v_{i,j}^{p,c} .
\end{align}
So far, this formulation for $M^{p,c}$ does not consider the direction of the power flow of $v_{i,j}^{p,c}$. 
However, it can make quite a difference since we only want to charge line usage in the direction of physical power flow. Suppose the pair power flow goes in the opposite direction. In that case, we have three possibilities: (a) bill the absolute value, (b) reimburse because the opposite direction means that the trade reliefs strain from the line, and (c) do not charge line costs. Preliminary experiments have shown that option (b) favors trades in opposing directions to the physical power flow, pushing the optimization away from the main objective. Option (a) leads to unfair allocations since trades are billed that effectively help stabilize the grid operation. 
Consequently, this proof-of-concept formulation settles with option (c).

Therefore, we redefine Eq.~\eqref{eq:cost-allocation-1} as follows
\begin{align}\label{eq:cost-allocation}
    M^{p,c} = \rho \sum_{(i,j) \in E} u_{i,j} \left[\mathrm{sign}(w_{i,j}) v_{i,j}^{p,c}\right]^+,
\end{align}
with $[\cdot]^+ = \max\{\cdot, 0\}$.
Equipped with the cost allocation for a single trade pair, we can split the costs between producer and consumer equally, considering all trades that they are involved in
\begin{align}
    M^i(x) = \frac{1}{2} \begin{cases}
        \sum_c x_{p,c} M^{p,c}& \text{if } i\in P\\
        \sum_p x_{p,c} M^{p,c}& \text{if } i\in C.
    \end{cases}
\end{align}

\subsection{Optimization Model}
Finally, we can combine the cost allocation with the power flow matching into a single objective by adding a matching penalty parameter $\alpha$ [$\text{ct}/\text{kWh}^2$]
\begin{align}\label{eq:qubo}
    \min_{x}\left[ \alpha \sum_{(i,j)\in E} \left(w_{i,j} - v_{i,j}(x)\right)^2 + \sum_{p,c} x_{p,c} M^{p,c}\right].
\end{align}
The problem is already in QUBO form, so no additional transformations are required to apply quantum methods like QA. The parameters $\rho$ and $\alpha$ must be chosen for the specific application instance. Technically, only a single parameter is necessary to define the optimization problem's outcome fully. However, using $\rho$ already in correct units for cost allocation makes handling the objective function and interpreting the results more straightforward.

\subsection{Final Tariffs}
After the optimization, we receive an optimal assignment $x^*$. Since we only perform discrete assignments, it is possible that some customers are not fully satisfied and trade more (or less) electricity than initially asked for (or offered), e.g., a producer aims to sell $-d^p$, but $\tilde{d}^p(x) = \sum_{c} d^{p,c}x^*_{p,c} < -d^p$. In contrast to the literature formulation in Ref.~\cite{baroche2019}, we consider the remaining volume to be sourced from the DSO. Therefore, the difference between required demand and achieved demand has to be bought (or sold) using flat tariffs, i.e., $\lambda^\text{buy} = 30 \text{ct}/\text{kWh}$ (and $\lambda^\text{sell} = 8 \text{ct}/\text{kWh})$. Furthermore, the consumer compensates the producer with a flat equilibrium tariff $\lambda^\text{eq} = (\lambda^\text{buy} + \lambda^\text{sell}) / 2 = 19 \text{ct}/\text{kWh}$ that is in between grid sell and buy price. Tariffs mentioned here are exemplary only and can be adapted without loss of generality. 

Combining these considerations, we can compute the final costs $\widetilde{M}$ or profits for producers and consumers separately:
\begin{align}\label{eq:producer_price}
    \widetilde{M}^p(x) = &M^p(x) + \lambda^\text{buy} \left[\tilde{d}^p(x) + d^p\right]^+\nonumber \\ & +\lambda^\text{sell} \left[\tilde{d}^{p}(x) + d^p\right]^- -\lambda^\text{eq} d^p(x)
\end{align}
\begin{align}\label{eq:consumer_price}
    \widetilde{M}^c(x) = &M^c(x) + \lambda^\text{buy} \left[d^c - \tilde{d}^c(x)\right]^+\nonumber \\ &+\lambda^\text{sell} \left[d^c - \tilde{d}^c(x)\right]^- + \lambda^\text{eq} d^p(x),
\end{align}
where $[\cdot]^- = \min\{\cdot, 0\}$. This ensures that excess/deficit energy amounts will be traded with the DSO.
Using the total demand of a peer, we can subsequently calculate the effective tariff per unit of electricity $\lambda^i = \widetilde{M}^i / d^i$.

\subsection{Possible Model Extensions}
There are a few considerations to make when applying the problem to real-world scenarios. A non-exhaustive list of example extensions is included below:

\paragraph{Grid connection} The problem described above is self-contained, i.e., we are in a self-reliant community. However, in principle, general problems can also be considered. This can be achieved by including the grid connection as an additional trade partner for all consumers and producers, increasing the number of binary variables by $|V|$.

\paragraph{Baseline Grid Fee} Grid costs are only determined by line usage. This may cause exaggerated price differences between participants. An additional small constant base grid fee can restrict these price differences.

\paragraph{Bidding Constraints} In literature, P2P markets are often realized by an auction market in which participants can bid a price for buying or selling their electricity \cite{Doan2021PeertoPeerET,Muhsen2022BusinessMO}. The current formulation does not involve the customer's action itself. An extension to the current scheme can be realized by employing constraints that disallow any assignments where the bids are violated, e.g., if a producer only sells for a certain amount, any assignment resulting in a profit lower than the respective amount will be disallowed by the constraints.

\section{Case Study}
\label{sec:case_study}

\begin{figure*}
    \centering
    \def\localscale{0.55}
    \includegraphics[scale=\localscale]{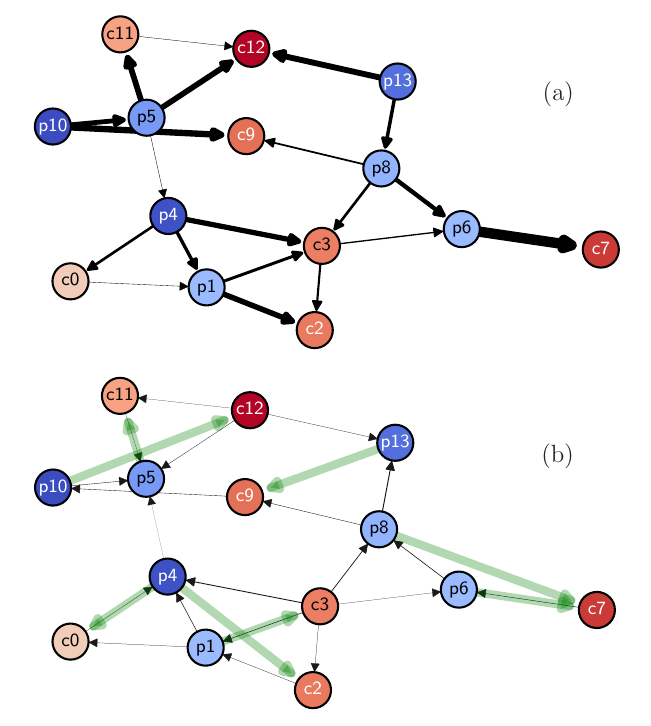}
    \includegraphics[scale=\localscale]{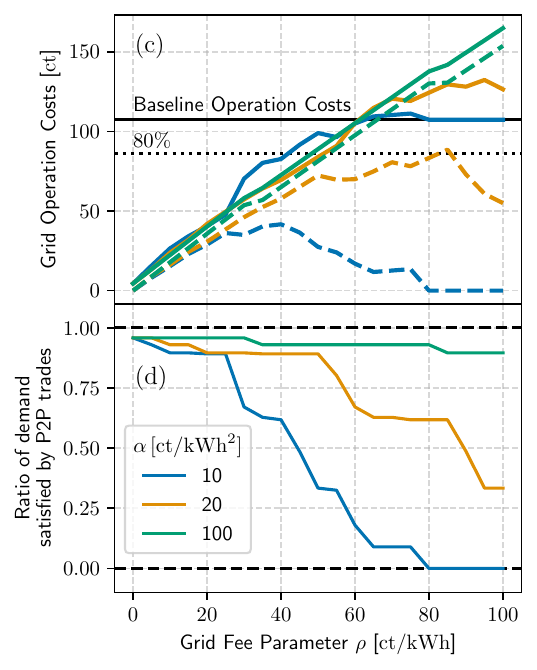}
    \includegraphics[scale=\localscale]{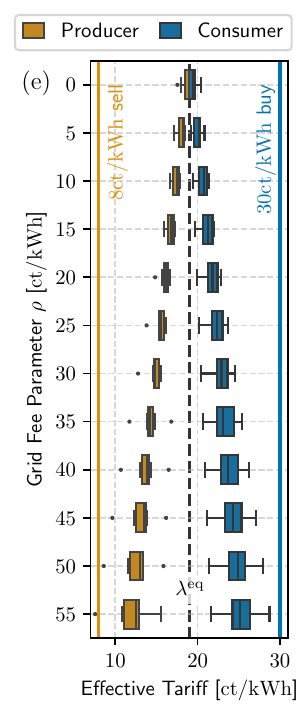}
    \caption{P2P power flow matching on an example instance of the IEEE \textsf{case14}. Panel~(a) shows the input data to the optimization problem consisting of the net topology, the consumer and producer data, as well as the physical power flow. Below, in~(b), we depict the solution to the optimization problem Eq.~\eqref{eq:qubo}; the green arrows indicate the matched pairs and the thickness of the power lines corresponds to the mismatch between physical and logical power flow. In the center,~(c) and~(d), the solution in terms of gathered grid operation costs is shown depending on the model parameters $\rho$ and $\alpha$. Finally,~(e) visualizes the effective tariffs in dependence of $\rho$ for the consumers and producers separately.}
    \label{fig:case_study}
\end{figure*}

Before focusing on the capabilities of quantum approaches for solving the P2P matching, we investigate how the model behaves under the parameters $\rho$ and $\alpha$. To that end, we examine the IEEE power system test case 14 (\textsf{case14}), provided by \texttt{pandapower}~\cite{thurner2018a} in the following. It consists of 14 buses and 20 power lines. Since we focus on residential P2P operation, we solely take the grid topology from the test case and rescale and replace the power lines with residential lines. The scaling factor is determined such that the shortest line is $50\,\text{m}$ long, the voltage level is dropped to $400\,\text{V}$, and NAYY
4x50 SE power lines are employed. We sample the production and consumption data for the customers from two shifted normal distributions centered around $\pm 1\,\text{kWh}$. Additionally, we rescale the production side so that the net consumption within the community becomes zero.
%The scheme for generating the problem instances is discussed in more detail in Ref.~\cite{bucher2024}.
A single instance of the problem is displayed in Fig.~\ref{fig:case_study}(a). The thickness and direction of the lines indicate the physical power flow obtained by \texttt{pandapower}'s DC power flow.

Fig.~\ref{fig:case_study}(b) shows the matches found by solving the model exactly with Gurobi using parameters $\rho = 45\,\ctpkwh$ and $\alpha = 100\,\ctpkwhs$. The thickness of the power lines indicates the mismatch between the physical and logical power flow here.

\subsection{Collected Grid Operation Fees}
In a non-P2P environment, electricity prices for customers do not directly reflect production costs. Besides taxes, a significant portion is attributed to operational costs for maintaining and stabilizing the power grid. In P2P markets, those costs still need to be covered as the market members use the existing infrastructure and incur maintenance costs. Let us assume that 50\% of the purchase price for electricity is the grid operation compound, i.e., $15\,\ctpkwh$ of the $30\,\ctpkwh$ baseline\footnote{Producers are not charged for grid operation in this assumption.}. The horizontal line in Fig.~\ref{fig:case_study}(c) represents the total collected grid operation fees of the DSO in the \textsf{case14} non-P2P scenario.

The dashed lines in Fig.~\ref{fig:case_study}(c) indicate the total cost allocation for different $\alpha$'s depending on the grid fee parameter $\rho$. As expected, the earnings from P2P trades for the DSO increase with increasing $\rho$; however, they start to fall off again as soon as costs become too expensive for the customers and they rather trade directly with the DSO again. This is also visualized in Fig.~\ref{fig:case_study}(d), where the ratio of P2P trades compared to total demand is shown. When customer demand is not satisfied by P2P trades, it will be saturated by direct DSO electricity purchases, enabling the DSO to subsequently collect additional grid operation fees even if no P2P trades are facilitated. The total collected grid operation fees are observable in the solid-line plot in panel (c). For $\alpha = 10\ctpkwhs$, we see that when no P2P trades occur, we precisely retrieve the collected baseline operation fees.

Furthermore, it is apparent that the total retrieved grid operation fees exhibit a similar trend up until they are saturated ($\rho < 60\ctpkwh$), independent of $\alpha$. We, therefore, require $\rho \approx 45 \ctpkwh$ to gain about 80\% of the baseline costs. The authors arbitrarily choose 80\%, which is sensible since the self-balancing community, in which P2P trades are facilitated, causes less strain on the power grid and, therefore, less operational costs.

\subsection{Individual Customer Tariffs}

Due to the different line usage, each customer will receive a different effective tariff composed of grid operation cost allocation and the compensation between consumer and producer. The final price or tariff is computed with Eqs.~\eqref{eq:producer_price} and~\eqref{eq:consumer_price}. Fig.~\ref{fig:case_study} (e) shows the different effective customer tariffs for consumers and producers for different grid fee parameters $\rho$. Clearly, at $\rho = 0$, all trades are free of charge. Hence, the consumer and producer tariffs align at the equilibrium tariff $\lambda^\text{eq} = 19\,\ctpkwh$. As the fee increases, consumption becomes more expensive and production less profitable. The average consumption tariff approaches the baseline tariff of $\lambda^\text{buy} = 30\,\ctpkwh$, while the consumer compensation tariff decreases to $\lambda^\text{sell} = 8\,\ctpkwh$. The experiments for the effective customer tariffs shown here have been conducted with $\alpha = 100\ctpkwhs$.

% Check the tense of all sentences: Presence preferred instead of past
% Use textsf for cases

\section{Benchmarks}\label{sec:benchmarks}

One aim of this study is to evaluate the applicability of quantum approaches to solving this problem. Therefore, we compare the performance of different solvers against each other in the following.

\subsection{Experimental Setup}
First, we explain the benchmarking setup, i.e., the considered problem instances, solvers, and metrics.
% paragraph transition
\subsubsection{Problem Instance Dataset} \label{sec:bencharmk_problems}
For testing on close-to real-world problems, we require grid topologies that resemble the future electricity grid, i.e., energy grids containing decentralized power generation. To this end, we modify the IEEE system test cases, as described in Sec.~\ref{sec:case_study}.
For our analysis, we select the available IEEE test cases between 9 and 57 nodes, referred to in the following as \textsf{case}\{\textsf{9}, \textsf{14}, \textsf{24}, \textsf{33}, \textsf{39}, \textsf{57}\} where the numbers denote the number of nodes in the respective IEEE power systems test case. For each test case, we generate 20 instances using different seeds to sample from the consumption distributions to generate different load profiles. 

Like Sec.~\ref{sec:case_study}, we determine $\rho$ for each test case, such that the DSO receives roughly 80\% of the baseline costs. Unsurprisingly, the grid topology significantly impacts the collected fees through cost allocation. Therefore, we must set case-specific grid fee constants, as seen in Table~\ref{table:rho/timeout}.

\subsubsection{Investigated Solvers}
We analyze the performance of classical branch-and-cut (Gurobi), classical heuristics (SA, TS), hybrid quantum-classical (Leap), and QA (D-Wave).

Specifically, we utilize D-Wave Advantage 5.4, located in Germany. Since the embedding of the fully connected optimization QUBO~\eqref{eq:qubo} onto the D-Wave hardware graph is limited to maximally 169 binary variables, \textsf{case33} with 272 binary variables is already too large for D-Wave. For embedding, we use the Clique embedder, a fast (polynomial-time) algorithm designed explicitly for fully connected problems~\cite{boothby2016}.

On the other hand, Gurobi can be faced with two equivalent problem formulations that make a difference in runtime and solution quality. As the QUBO~\eqref{eq:qubo} has a mean-squared-error form, we utilize \texttt{CVXPY}\cite{diamond2016} to interface Gurobi.
\texttt{CVXPY} formulates the optimization problem as follows
\begin{align}\label{eq:convex}
    \min_{x} &\alpha \sum_{i, j}\zeta_{i,j}^2 + \sum_{p,c} x_{p,c} M^{p,c}\nonumber  \\
    \text{s.t.} \quad & \zeta_{i,j} = w_{i,j} - v_{i,j}(x) \quad \forall i,
\end{align}
where $\zeta_{i,j}$ is the residual power flow per line. This formulation allows Gurboi to find better lower bounds more efficiently, leading to a faster converging branch-and-cut procedure. The second alternative is to present Gurobi the expanded expression of Eq.~\eqref{eq:qubo}, which is referred to in the following as Gurobi[ncvx].

Since the solution bit-string will only be sparsely populated with ones, i.e., only a tiny fraction of all possible trades will be satisfied, we initialize the SA solver with all bits set to zero. This will make the exploration stage easier at the beginning since SA requires a few bitflips to arrive at a viable solution.

To ensure a fair comparison, we optimize the hyperparameters of the classical heuristics and the D-Wave QA. The results can be found in Table~\ref{table:optimal_hyperparameters}, and further details are presented in Appendix~\ref{sec:optimal_hyperparameters}.

All classical experiments were conducted on a single core of an AMD Ryzen Threadripper PRO 5965WX.

\begin{table}
    \small
    \centering
    \caption{$\rho$ values and timeout settings for each case}
    \begin{center}
        \begin{tabular}{l | c c c c c c}
        \textsf{case} & \textsf{9} & \textsf{14} & \textsf{24} & \textsf{33} & \textsf{39} & \textsf{57} \\ \hline
        $\rho\,[\ctpkwh]$ & 20 & 45 & 40 & 15 & 15 & 15 \\
        Timeout$\,[\text{s}]$ & 1 & 2.45 & 7.2 & 13.6 & 19 & 40.6
        \end{tabular}
    \end{center}
    \label{table:rho/timeout}
\end{table}

\subsubsection{Benchmarking Strategies}

In the following, we devise two benchmarking strategies to compare the solvers on solution quality and runtime.

\paragraph{Solution quality within time limit} For this strategy, we set a fixed timeout for each test case and investigate how close the solutions are to the optimal solution. The reference solution $x^*$ is obtained by running Gurobi for one hour for each instance using the formulation created with \texttt{CVXPY}~\cite{diamond2016}. The relative objective error can then be computed using
\begin{align}\label{eq:metric-relerr}
    \varepsilon(x) = \frac{C(x) - C(x^*)}{C(x^*)},
\end{align}
where $C(x)$ is the cost function defined in Eq.~\eqref{eq:qubo}.

We choose a linearly growing timeout with the number of involved binary variables to mitigate the growing problem difficulty with increasing problem size. We set the timeout for each use case according to the number of possible trades, i.e., $|P \times C|$, or approximately $(N/2)^2$, which is subsequently used. We arbitrarily set \textsf{case9} to 1$\,\text{s}$, and scale the remaining cases accordingly, see Table~\ref{table:rho/timeout}. Leap has a minimal configurable time limit of 3$\,\text{s}$, which overrides our timeout setting.

% The embedding onto D-Wave's hardware graph drastically limits the problem size that QA can solve. \textsf{Case24}, with 144 logical qubits, already requires about 2000 of the more than 5000 available physical qubits. Maximally problems consisting of 169 logical qubits are embeddable.

\paragraph{Time to Solution (TTS)}
As we are investigating some nondeterministic heuristics with varying success probability, it is sensible to inspect the expected time each solver requires to reach a solution of adequate quality. This evaluation considers solutions acceptable if they are within $5\%$ of the optimal solution computed using Gurobi ($\varepsilon \leq 5 \%$). Furthermore, we only run experiments until \textsf{case33}, as the optimal solution has not been found with Gurobi for the larger cases, possibly distorting the TTS. The error threshold is chosen since some solvers struggle to find the optimal solution, disallowing TTS calculation. We run each solver on each instance until 50 samples have been found that are within $5\%$ of the optimal solution. However, since this might take a very long time, we fixed an upper limit of 1000\,s. To ensure the statistical significance of our results, we disregard the TTS if we do not find at least 10 samples within the timeout. 

TTS is computed by estimating the expected runtime based on the sample time $t_s$ and the expected number of repetitions to sample one solution with 99\% probability, based on the probability measured from the samples $p_\varepsilon$~\cite{steiger2015}
\begin{align}\label{eq:metric-tts}
    \text{TTS}(\varepsilon) = t_s \left\lceil \frac{\log(1 - 0.99)}{\log(1 - p_{\varepsilon})} \right\rceil.
\end{align}

Since Leap only has the time limit as a controllable parameter, we cannot directly evaluate TTS here. Instead, we investigate how a growing time limit improves solution quality.

\subsection{Results}

\begin{figure}
    \centering
    \includegraphics[width=\linewidth]{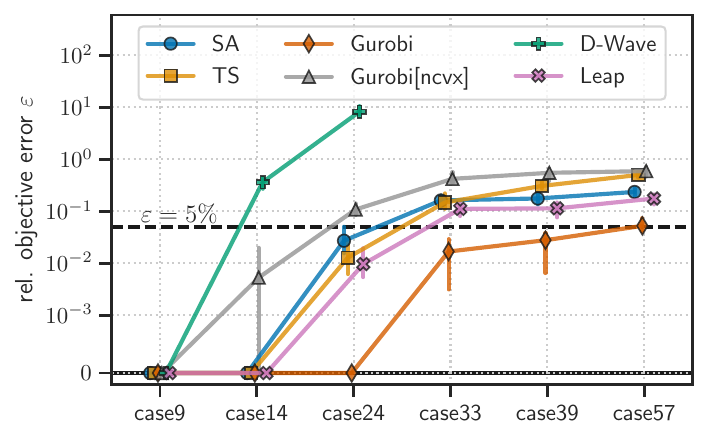}%
    \caption{The median relative objective error for different solvers and test cases. The error bars mark the 50\% percentile interval. Gurobi's median solution quality stays below the 5\% rel. error mark for almost all cases.}
    \label{fig:timeout}
\end{figure}

\begin{figure}
    \centering
    \includegraphics[width=\linewidth]{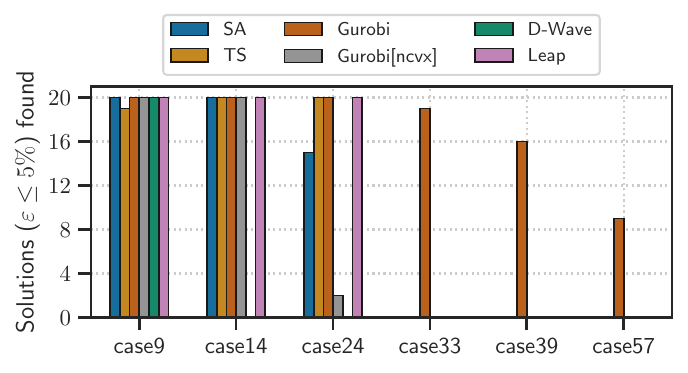}%
    \caption{Number of instances for which a solution within the 5\% error bound has been obtained.}
    \label{fig:num_close_to_optimum_results}
\end{figure}

\paragraph{Solution Quality}
Fig.~\ref{fig:timeout} shows the relative error with the time limit from Table~\ref{table:rho/timeout} for different cases. We can see that, except for D-Wave and the Gurobi[ncvx], almost all algorithms solve the first two cases to optimality. Afterward, we see a steep decrease in solution quality for every solver, crossing the $5\%$ error threshold between \textsf{case24} and \textsf{case33} except for the convex Gurobi, which exhibits the lowest error ($< 5\%$). This is also highlighted in Fig. \ref{fig:num_close_to_optimum_results}, where Gurobi is the only algorithm that consistently finds solutions that are within 5\% of the optimal solution for every case. SA, TS, and Leap perform similarly, with Leap having a slight advantage. D-Wave performs very poorly except for the smallest case (20 binary variables). It exhibits a mean relative error of more than $20\%$ in \textsf{case14} and almost $1000\%$ in \textsf{case24}.

Notably, Gurobi[ncvx] performs considerably worse than Gurobi and possesses higher solution quality errors than the classical heuristics and the hybrid quantum algorithm.

\begin{figure}
    \centering
    \includegraphics[width=\linewidth]{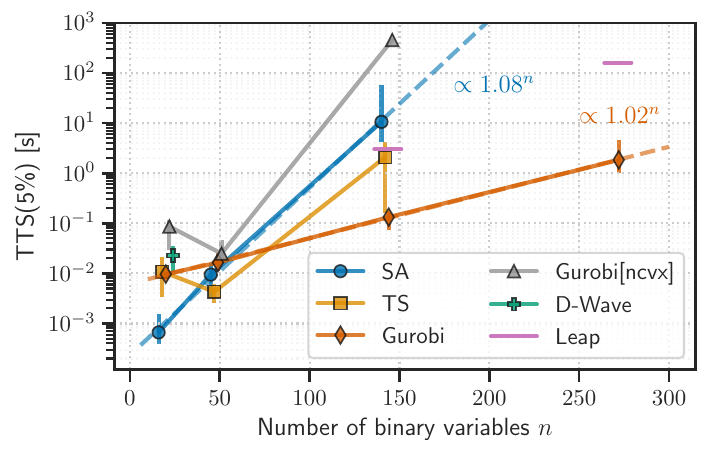}%
    \caption{Median TTS to find a solution that is within $5\%$ of the optimal solution with respect to the number of binary variables in the optimization problem. Error bars indicate the 50\% percentile interval. D-Wave only shows a single point since no solutions within the error bound were observed for problem instances larger than \textsf{case9}.}
    \label{fig:TTS}
\end{figure}

\paragraph{Time to Solution}
Fig.~\ref{fig:TTS} presents the estimated TTS results to reach solutions within the 5\% error. We are missing several data points because specific solvers could not find the minimum required amount of acceptable samples within the timeout limit, as discussed in the following. Due to the solution quality results, we did not attempt to find TTS estimates with D-Wave for cases larger than \textsf{case9}. Similarly, SA, TS, Gurobi[ncvx] do not find sufficient satisfactory solutions for \textsf{case33}, where they crossed the error 5\% error threshold in Fig.~\ref{fig:timeout}, even with the increased time limit considered.

Both Gurobi and SA exhibit exponential runtime scaling, as the linear dependence with the logarithmic time scale is distinctly visible. A linear regression reveals exponential scaling with $\propto 1.08^n$ for SA and $\propto 1.02^n$ for Gurobi, with $R^2$ scores above $99\%$, where $n$ is the number of binary variables in the optimization problem. Interestingly, TS and Gurobi[ncvx] both experience a similar dip in required runtime between \textsf{case9} and \textsf{case14}, indicating that \textsf{case14} could be easier than \textsf{case9}. However, this hypothesis is not supported by the results from SA and Gurobi, indicating a similarity between the solving strategy of Gurobi[ncvx] and TS. Since the three data points of TS and Gurobi[ncvx] do not exhibit apparent exponential scaling, we do not perform linear regression here. However, qualitatively, TS seems to scale similarly to SA, and Gurobi[ncvx] scales worse than SA.

Focusing on the absolute TTS at the smallest \textsf{case9}, we see that SA is fastest ($< 1\,\mathrm{ms}$), Gurobi and TS come second with $\approx 10\,\mathrm{ms}$. Afterwards, we identify D-Wave ($\approx 20\,\mathrm{ms}$) and Gurobi[ncvx] ($\approx 100\,\mathrm{ms}$) as the slowest.
% The TTS to reach a solution within $5\%$ of the optimal solution remains fairly constant between \textsf{case9} and \textsf{case14}. Afterward, we see a steep increase in TTS for every solver. SA initially performs best on \textsf{case9} and subsequently performs worst on \textsf{case24}. The spread of TTS is a lot bigger for the metaheuristics SA and TS than for Gurobi. Again, Gurobi outperformed the other algorithms on the larger instances, being the only algorithm expected to find an acceptable solution in a reasonable time for \textsf{case33}. 

\begin{figure}
    \centering
    \includegraphics[width=\linewidth]{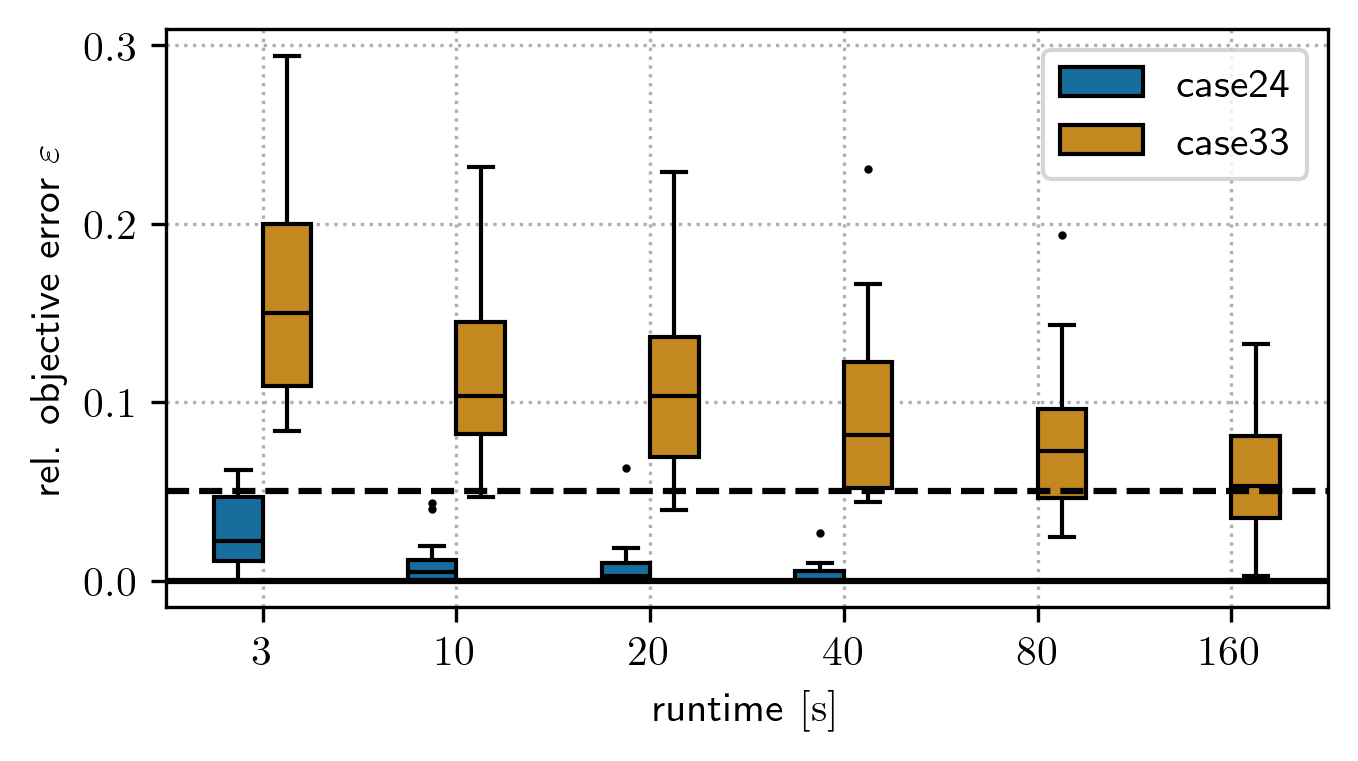}%
    \caption{D-Wave hybrid Leap BQM solver solution quality improvement against timeout setting.}
    \label{fig:leap}
\end{figure}

Since TTS analysis is impossible with D-Wave's Leap hybrid solver, we separately analyze the solution quality improvement upon growing time limit settings in Fig.~\ref{fig:leap}. Due to the minimum timeout of 3\,s for Leap, we cannot infer the TTS for \textsf{case24} as it is certainly below 3\,s. For \textsf{case33}, we can estimate that the TTS is probably larger than 160\,s, since the median point crosses the 5\% threshold there. As a consequence, the median TTS is about 160\,s, which is slower than Gurobi but probably faster than SA, judging from Fig.~\ref{fig:TTS}, where 3\,s and 160\,s are indicated with faint pink horizontal lines.

\subsection{Discussion}
The results show that Gurobi---equipped with the convex problem formulation---outperforms (or performs on par with) every other solver in terms of solution quality within a time limit. Only the TTS of SA and TS is faster for the small test cases \textsf{case9} and \textsf{case14}, as apparent from Fig.~\ref{fig:TTS}, which can be explained by the fast sampling times of heuristics in these small instances and initialization overhead of the more complex Gurobi solver.

That Gurobi outperforms the classical and quantum-classical heuristics is to some extent surprising since the problem of Eq.~\eqref{eq:qubo} is inherently of unconstrained form, which means no constraints had to be formulated as penalty, which typically makes problems more challenging to solve with unconstrained solving algorithms~\cite{lucas2014b}. The analysis of both formulations for Gurobi unmistakably showed that presenting Gurobi with the convex formulation is advantageous. Our results on the scaling behavior of the TTS of Gurboi and SA also show superior scaling of Gurobi compared to SA, indicating that no break-even point can be expected with growing problem size.

Furthermore, QA performs extremely sub-par in this use case, already losing solution quality beyond the accepted threshold of 5\% relative error in the second smallest test case. This result can be explained by the embedding necessary to map the fully connected problem structure onto the graph of the D-Wave hardware. Even though we utilize the specialized Clique embedding method for fully connected problems with chain strength hyperparameter optimization, the embedding still requires excessive amounts of physical qubits for a single logical one.

The findings from this specific problem of power flow matching tentatively suggest that binary least-squares problems, or even binary convex problems, might not be suitable candidates for quantum optimization applications, as they can be more efficiently solved using existing classical optimization algorithms. However, more extensive studies are needed to prove the generalization of this statement. It is important to remark here that quantum specialized optimization methods for convex problems~\cite{abbas2023} have not been investigated within that study but only QUBO optimization methods like QA. Nevertheless, these oracle-based methods are expected only to apply when fault-tolerant QC is available.
\section{Conclusion}\label{sec:conclusion}

This study had two objectives: first, to develop an optimization model for allocating grid usage costs in P2P electricity markets, and second, to investigate whether quantum optimization techniques can successfully be employed to solve the defined model.

For the first part, we devised a binary optimization model that aims to match the logical power flow---the power flow that can be attributed to P2P consumer-producer pairs---with the physical power flow when all producers and consumers in the network inject (draw) power into (from) the grid. An in-depth investigation of the model behavior showed that this formulation could retrieve all collected grid operation fees from beforehand. This indicates that the formulation is viable for attributing P2P grid operation costs. Nevertheless, the formulation is still rudimentary and probably requires extensions (e.g.,  external grid connection, baseline grid fees, and bidding constraints) to make it applicable in a real-world scenario. Yet, it serves as a sufficient proof-of-concept use case for benchmarking and a seed for future work.

The second goal is to show the applicability of quantum optimization techniques. This is straightforwardly possible as the problem formulation is already in QUBO form, something that can be atypical in instances of applied quantum computing to real-world use cases. Thus, we directly benchmarked QA against classical heuristics, hybrid quantum-classical heuristics, and a classical branch-and-cut solver. When posed with a convex formulation of the least-squares problem, the classical solver outperformed any heuristic, with QA performing especially sub-par due to the fully connected nature of the problem at hand. This significantly challenges the useful application of QA to solve the power flow matching problem, as classical methods find better solutions faster and exhibit more favorable scaling behavior. Furthermore, these results indicate that binary least-squares problems can be inherently challenging for quantum methods and classical heuristics. Whether this consequence extends to general convex optimization problems remains an open question that requires further investigation. Nevertheless, specialized convex quantum optimization algorithms such as Refs.~\cite{apeldoorn2020,chakrabarti2020,abbas2023} may be a more advanced future route for solving the power flow matching problem.

\section*{ACKNOWLEDGEMENTS}
The authors would like to thank Kumar Ghosh and Naeimeh Mohsehni for their helpful discussions throughout this research.
This work was supported by the German Federal Ministry of Education and Research under the funding program ``Förderprogramm
Quantentechnologien – von den Grundlagen zum Markt'' (funding program quantum technologies — from basic research to market), project
Q-Grid, 13N16177.

\bibliographystyle{apalike}
{\small
\bibliography{references}}

\begin{thebibliography}{}

\bibitem[Abbas et~al., 2024]{abbas2023}
Abbas, A., Ambainis, A., Augustino, B., B{\"a}rtschi, A., Buhrman, H., Coffrin, C., Cortiana, G., Dunjko, V., Egger, D.~J., Elmegreen, B.~G., Franco, N., Fratini, F., Fuller, B., Gacon, J., Gonciulea, C., Gribling, S., Gupta, S., Hadfield, S., Heese, R., Kircher, G., Kleinert, T., Koch, T., Korpas, G., Lenk, S., Marecek, J., Markov, V., Mazzola, G., Mensa, S., Mohseni, N., Nannicini, G., O'Meara, C., Tapia, E.~P., Pokutta, S., Proissl, M., Rebentrost, P., Sahin, E., Symons, B. C.~B., Tornow, S., Valls, V., Woerner, S., {Wolf-Bauwens}, M.~L., Yard, J., Yarkoni, S., Zechiel, D., Zhuk, S., and Zoufal, C. (2024).
\newblock Challenges and opportunities in quantum optimization.
\newblock {\em Nature Reviews Physics}, pages 1--18.

\bibitem[Baroche et~al., 2019]{baroche2019}
Baroche, T., Pinson, P., Latimier, R. L.~G., and Ahmed, H.~B. (2019).
\newblock Exogenous {{Cost Allocation}} in {{Peer-to-Peer Electricity Markets}}.
\newblock {\em IEEE Transactions on Power Systems}, 34(4):2553--2564.

\bibitem[Blenninger et~al., 2024]{blenninger2024}
Blenninger, J., Bucher, D., Cortiana, G., Ghosh, K., Mohseni, N., N{\"u}{\ss}lein, J., O'Meara, C., Porawski, D., and Wimmer, B. (2024).
\newblock Q-{{GRID}}: {{Quantum Optimization}} for the {{Future Energy Grid}}.
\newblock {\em KI - K{\"u}nstliche Intelligenz}.

\bibitem[Boothby et~al., 2016]{boothby2016}
Boothby, T., King, A.~D., and Roy, A. (2016).
\newblock Fast clique minor generation in {{Chimera}} qubit connectivity graphs.
\newblock {\em Quantum Information Processing}, 15(1):495--508.

\bibitem[Born and Fock, 1928]{born1928}
Born, M. and Fock, V. (1928).
\newblock {Beweis des Adiabatensatzes}.
\newblock {\em Zeitschrift f{\"u}r Physik}, 51(3):165--180.

\bibitem[Bucher et~al., 2024a]{bucher2024a}
Bucher, D., Kraus, N., Blenninger, J., Lachner, M., Stein, J., and {Linnhoff-Popien}, C. (2024a).
\newblock Towards {{Robust Benchmarking}} of {{Quantum Optimization Algorithms}}.

\bibitem[Bucher et~al., 2024b]{bucher2024}
Bucher, D., Porawski, D., Wimmer, B., N{\"u}{\ss}lein, J., O'Meara, C., Mohseni, N., Cortiana, G., and {Linnhoff-Popien}, C. (2024b).
\newblock Evaluating {{Quantum Optimization}} for {{Dynamic Self-Reliant Community Detection}}.

\bibitem[Chakrabarti et~al., 2020]{chakrabarti2020}
Chakrabarti, S., Childs, A.~M., Li, T., and Wu, X. (2020).
\newblock Quantum algorithms and lower bounds for convex optimization.
\newblock {\em Quantum}, 4:221.

\bibitem[Christie et~al., 2000]{christie2000}
Christie, R., Wollenberg, B., and Wangensteen, I. (2000).
\newblock Transmission management in the deregulated environment.
\newblock {\em Proceedings of the IEEE}, 88(2):170--195.

\bibitem[Colucci et~al., 2023]{colucci2023}
Colucci, G., Linde, S. V.~D., and Phillipson, F. (2023).
\newblock Power {{Network Optimization}}: {{A Quantum Approach}}.
\newblock {\em IEEE Access}, 11:98926--98938.

\bibitem[Diamond and Boyd, 2016]{diamond2016}
Diamond, S. and Boyd, S. (2016).
\newblock {{CVXPY}}: {{A Python-Embedded Modeling Language}} for {{Convex Optimization}}.
\newblock https://arxiv.org/abs/1603.00943v2.

\bibitem[Doan et~al., 2021]{Doan2021PeertoPeerET}
Doan, H.~T., Cho, J., and Kim, D. (2021).
\newblock Peer-to-peer energy trading in smart grid through blockchain: A double auction-based game theoretic approach.
\newblock {\em IEEE Access}, 9:49206--49218.

\bibitem[Farhi et~al., 2001]{farhi2001}
Farhi, E., Goldstone, J., Gutmann, S., Lapan, J., Lundgren, A., and Preda, D. (2001).
\newblock A {{Quantum Adiabatic Evolution Algorithm Applied}} to {{Random Instances}} of an {{NP-Complete Problem}}.
\newblock {\em Science}, 292(5516):472--475.

\bibitem[{Fern{\'a}ndez-Campoamor} et~al., 2021]{fernandez-campoamor2021c}
{Fern{\'a}ndez-Campoamor}, M., O'Meara, C., Cortiana, G., Peric, V., and {Bernab{\'e}-Moreno}, J. (2021).
\newblock Community {{Detection}} in {{Electrical Grids Using Quantum Annealing}}.

\bibitem[Glover, 1989]{glover1989}
Glover, F. (1989).
\newblock Tabu {{Search}}---{{Part I}}.
\newblock {\em ORSA Journal on Computing}, 1(3):190--206.

\bibitem[{Gurobi Optimization}, 2024]{gurobi}
{Gurobi Optimization} (2024).
\newblock {Gurobi Optimizer Reference Manual}.

\bibitem[Kea et~al., 2023]{kea2023a}
Kea, K., Huot, C., and Han, Y. (2023).
\newblock Leveraging {{Knapsack QAOA Approach}} for {{Optimal Electric Vehicle Charging}}.
\newblock {\em IEEE Access}, 11:109964--109973.

\bibitem[Kirkpatrick et~al., 1983]{kirkpatrick1983a}
Kirkpatrick, S., Gelatt, C.~D., and Vecchi, M.~P. (1983).
\newblock Optimization by {{Simulated Annealing}}.
\newblock {\em Science}, 220(4598):671--680.

\bibitem[Le~Cadre et~al., 2020]{lecadre2020}
Le~Cadre, H., Jacquot, P., Wan, C., and Alasseur, C. (2020).
\newblock Peer-to-peer electricity market analysis: {{From}} variational to {{Generalized Nash Equilibrium}}.
\newblock {\em European Journal of Operational Research}, 282(2):753--771.

\bibitem[Lucas, 2014]{lucas2014b}
Lucas, A. (2014).
\newblock Ising formulations of many {{NP}} problems.
\newblock {\em Frontiers in Physics}, 2.

\bibitem[Mastroianni et~al., 2024]{mastroianni2024a}
Mastroianni, C., Plastina, F., Scarcello, L., Settino, J., and Vinci, A. (2024).
\newblock Assessing {{Quantum Computing Performance}} for {{Energy Optimization}} in a {{Prosumer Community}}.
\newblock {\em IEEE Transactions on Smart Grid}, 15(1):444--456.

\bibitem[McGeoch and Farr{\'e}, 2021]{mcgeochb}
McGeoch, C. and Farr{\'e}, P. (2021).
\newblock The {{Advantage System}}: {{Performance Update}}.
\newblock Technical report.

\bibitem[Mohseni et~al., 2024]{mohseni2024}
Mohseni, N., Morstyn, T., Meara, C.~O., Bucher, D., N{\"u}{\ss}lein, J., and Cortiana, G. (2024).
\newblock A {{Competitive Showcase}} of {{Quantum}} versus {{Classical Algorithms}} in {{Energy Coalition Formation}}.

\bibitem[Muhsen et~al., 2022]{Muhsen2022BusinessMO}
Muhsen, H., Allahham, A., Al-Halhouli, A.~T., Al-Mahmodi, M., Alkhraibat, A., and Hamdan, M. (2022).
\newblock Business model of peer-to-peer energy trading: A review of literature.
\newblock {\em Sustainability}.

\bibitem[Nielsen and Chuang, 2010]{nielsen2010}
Nielsen, M.~A. and Chuang, I.~L. (2010).
\newblock {\em Quantum Computation and Quantum Information}.
\newblock Cambridge university press, Cambridge, 10th anniversary edition edition.

\bibitem[O'Meara et~al., 2023]{omeara2023}
O'Meara, C., {Fern{\'a}ndez-Campoamor}, M., Cortiana, G., and {Bernab{\'e}-Moreno}, J. (2023).
\newblock Quantum {{Software Architecture Blueprints}} for the {{Cloud}}: {{Overview}} and {{Application}} to {{Peer-2-Peer Energy Trading}}.
\newblock In {\em 2023 {{IEEE Conference}} on {{Technologies}} for {{Sustainability}} ({{SusTech}})}, pages 191--198.

\bibitem[Rajak et~al., 2023]{rajak2023}
Rajak, A., Suzuki, S., Dutta, A., and Chakrabarti, B.~K. (2023).
\newblock Quantum {{Annealing}}: {{An Overview}}.
\newblock {\em Philosophical Transactions of the Royal Society A: Mathematical, Physical and Engineering Sciences}, 381(2241):20210417.

\bibitem[R{\o}nnow et~al., 2014]{ronnow2014}
R{\o}nnow, T.~F., Wang, Z., Job, J., Boixo, S., Isakov, S.~V., Wecker, D., Martinis, J.~M., Lidar, D.~A., and Troyer, M. (2014).
\newblock Defining and detecting quantum speedup.
\newblock {\em Science}, 345(6195):420--424.

\bibitem[Sousa et~al., 2019]{sousa2019}
Sousa, T., Soares, T., Pinson, P., Moret, F., Baroche, T., and Sorin, E. (2019).
\newblock Peer-to-peer and community-based markets: {{A}} comprehensive review.
\newblock {\em Renewable and Sustainable Energy Reviews}, 104:367--378.

\bibitem[Steiger et~al., 2015]{steiger2015}
Steiger, D.~S., R{\o}nnow, T.~F., and Troyer, M. (2015).
\newblock Heavy {{Tails}} in the {{Distribution}} of {{Time}} to {{Solution}} for {{Classical}} and {{Quantum Annealing}}.
\newblock {\em Physical Review Letters}, 115(23):230501.

\bibitem[Thurner et~al., 2018]{thurner2018a}
Thurner, L., Scheidler, A., Sch{\"a}fer, F., Menke, J.-H., Dollichon, J., Meier, F., Meinecke, S., and Braun, M. (2018).
\newblock Pandapower---{{An Open-Source Python Tool}} for {{Convenient Modeling}}, {{Analysis}}, and {{Optimization}} of {{Electric Power Systems}}.
\newblock {\em IEEE Transactions on Power Systems}, 33(6):6510--6521.

\bibitem[van Apeldoorn et~al., 2020]{apeldoorn2020}
van Apeldoorn, J., Gily{\'e}n, A., Gribling, S., and de~Wolf, R. (2020).
\newblock Convex optimization using quantum oracles.
\newblock {\em Quantum}, 4:220.

\end{thebibliography}

\section*{APPENDIX}
\section{Hyperparameter Optimization} \label{sec:optimal_hyperparameters}

\begin{table}
    \centering
    \small
    \setlength{\tabcolsep}{8pt}
    \caption{Optimal hyperparameters for D-Wave}
    \begin{center}
        \begin{tabular}{l | c c c}
        \textsf{case} & \textsf{9} & \textsf{14} & \textsf{24}\\
        \hline\rule{0pt}{1.1\normalbaselineskip}%
        \texttt{chain\_strength\_factor} & 0.3 & 0.4 & 0.9 \\
        \texttt{annealing\_time} [$\mu\mathrm{s}$] & 10 & 20 & 20\\
        \end{tabular}
    \end{center}
    \label{table:optimal_hyperparameters}
\end{table}

% \begin{table*}
%     \centering
%     \small
%     \setlength{\tabcolsep}{8pt}
%     \caption{Optimal hyperparameters split between problem size and solver}
%     \begin{center}
%         \begin{tabular}{l | l | c c c c c c}
%         Solver & Parameter & \textsf{case9} & \textsf{case14} & \textsf{case24} & \textsf{case33} & \textsf{case39} & \textsf{case57}\\
%         \hline\rule{0pt}{1.1\normalbaselineskip}%
%         \multirow{3}{*}{SA} & \texttt{num\_sweeps} & 100 & 100 & 1000 & 5000 & 5000 & 10000\\
%         & \texttt{beta\_start} & 0.2 & 0.2 & 0.2 & 0.2 & 0.2 & 0.2 \\
%         & \texttt{beta\_end} & 1000 & 1000 & 1000 & 1000 & 1000 & 1000  \\
%         \hline\rule{0pt}{1.1\normalbaselineskip}%
%         \multirow{2}{*}{D-Wave} & \texttt{chain\_strength\_factor} & 0.3 & 0.4 & 0.9 & -- & -- & --\\
%         & \texttt{annealing\_time} [$\mu\mathrm{s}$] & 10 & 20 & 20 & -- & -- & -- \\
%         \end{tabular}
%     \end{center}
%     \label{table:optimal_hyperparameters}
% \end{table*}

To ensure fair benchmark comparison, we aim to devote equal effort to hyperparameter optimization of the individual solvers~\cite{bucher2024a}.
Leap cannot be fine-tuned, and Gurobi, as a commercial solver, is also considered with default parameters.
%\footnote{Gurobi chooses and adapts its parameters dynamically.}
Preliminary experiments showed little to no response of TS to changing hyperparameters; hence, we do not consider it in the following section. Instead, we only consider SA and D-Wave in the following, whose results are summarized in Table~\ref{table:optimal_hyperparameters}. 

\subsection{Simulated Annealing}

\begin{figure*}
    \centering
    \includegraphics[scale=0.53]{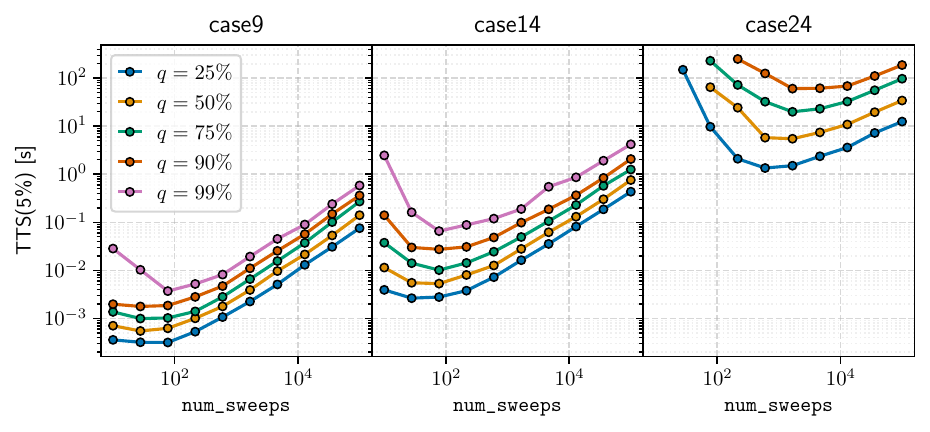}
    \includegraphics[scale=0.53]{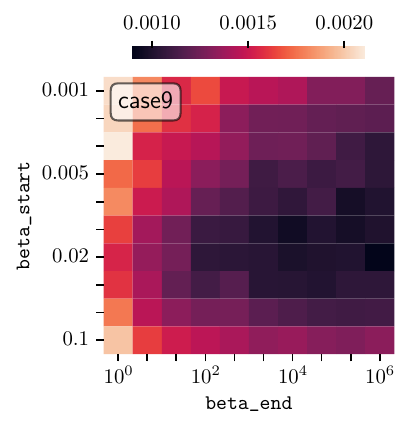}
    \includegraphics[scale=0.53]{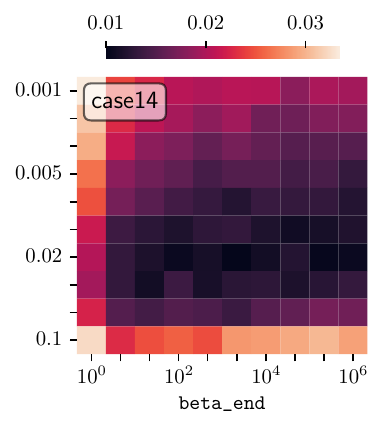}
    \caption{Left-hand side shows the TTS of SA for different \textsf{num\_sweeps} with annealing schedule $(0.02, 1000)$ showcasing different quantiles. The two heatmaps on the right show the 75\% quantile TTS depending on different \texttt{beta\_start} and \texttt{beta\_end} parameters of the annealing schedule for \textsf{case9} and \textsf{case14} at 100 Monte Carlo sweeps.}
    \label{fig:sa-hyperparameters}
\end{figure*}

The most critical hyperparameter of SA is the number of Monte Carlo sweeps (\texttt{num\_sweeps}) computed for a single sample. A larger number of sweeps results in better solutions, but the runtime for a single sample of the algorithm is directly proportional to the number of sweeps in the algorithm. Therefore, we expect a TTS sweet spot when tuning \texttt{num\_sweeps}, similar to Refs.~\cite{ronnow2014,steiger2015}. Indeed, Fig.~\ref{fig:sa-hyperparameters} shows that the smaller instances exhibit an optimal TTS of around 100 sweeps, while \textsf{case24} is optimal with around 1000 sweeps. For the larger test cases, it was not really possible to estimate the TTS correctly, as the samples were consistently above the 5\% error threshold. Hence, we heuristically set increasing \texttt{num\_sweeps} for these instances: 5000 sweeps for \textsf{case33} and \textsf{case39} and 10000 for \textsf{case57}.

The second set of parameters investigated is the start- and end-point of the geometric annealing schedule of SA. Since we have a fixed starting point, where all bits are set to zero, we conjecture that the optimal annealing schedule stays similar throughout different problem sizes. The right two plots of Fig.~\ref{fig:sa-hyperparameters} show combinations of different start and end values as a heatmap of \textsf{case9} and \textsf{case14}. Visually, we observe a similar large region with good TTS values for both cases, centered around $(0.2, 1000)$. Since the high computational demand required for this particular investigation, we only conducted it on \texttt{case9} and \texttt{case14}.
Nevertheless, we tested the $(0.2, 1000)$ parameter setting against the default (automatically calculated) schedule for all instances and ultimately received better results averaged over all instances. Hence, we argue that this setting is overall beneficial, even if it may not be the best one for a single instance.

\subsection{D-Wave}

\begin{figure*}
    \centering
    \includegraphics[scale=0.57]{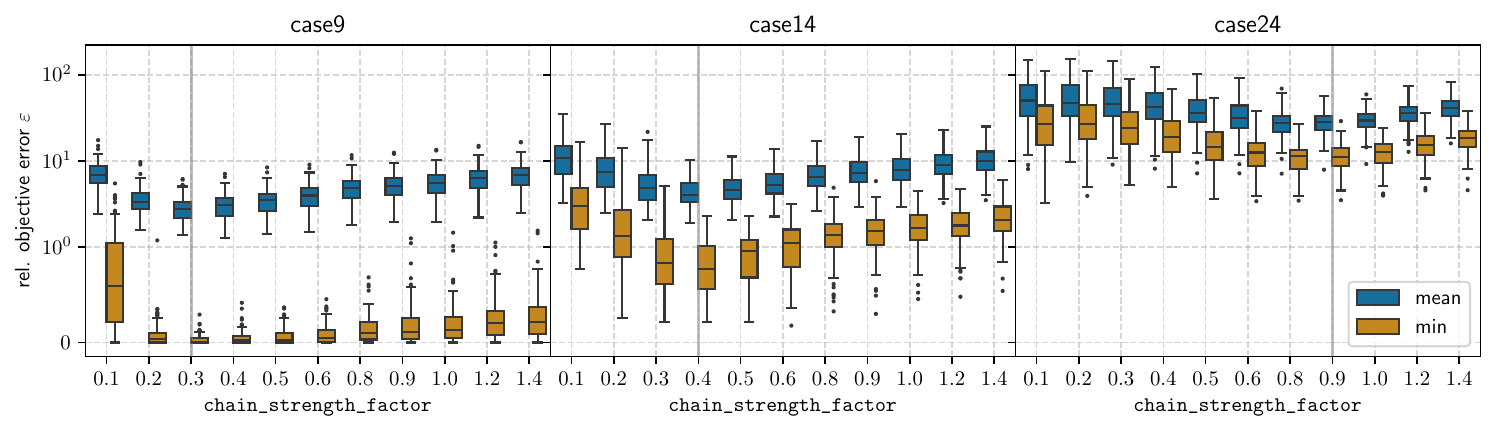}
    \caption{Average and Minimum relative energy error of D-Wave Advantage 5.4 with different \textsf{chain\_strength\_factor}s. Gray vertical lines mark the optimal \texttt{chain\_strength\_factor}.}
    \label{fig:dw-hyperparameters}
\end{figure*}

The most prominent parameter for QA is the \texttt{annealing\_time}, governing the overall time of the annealing process. In theory, a longer annealing time should result in better solution quality. However, due to the error-prone NISQ hardware, longer times most often introduce too much error, resulting in unusable samples. Nevertheless, our experiments showed little to no effect of the annealing time on the solution quality. Only a slight indication was found that for \textsf{case9} 10\,$\mu\text{s}$ is optimal, while for the remaining cases, the default 20\,$\mu\text{s}$ worked best.

As explained in the main text, embedding the fully connected QUBO problem onto the D-Wave QPU hardware graph requires the aggregation of multiple physical qubits to a single logical qubit to replicate the connectedness required for the input problem. Technically, this is done by augmenting the optimization problem with large biases between qubits belonging to the same binary variable. The strength of this interaction is commonly calculated using D-Wave's \texttt{uniform\_torque\_compensation}. However, it is heuristically computed from the biases in the input problem, so the strength setting may not be optimal. Therefore, we scale the bias using a \texttt{chain\_strength\_factor} that weakens the interactions if less than 1 and strengthens them otherwise.

Fig.~\ref{fig:dw-hyperparameters} shows the minimum and mean relative energy error from 1000 samples obtained from the D-Wave Advantage 5.4 machine. In all three embeddable cases, we can clearly observe a minimum, indicating the optimal \texttt{chain\_strength\_factor}, summarized in Table~\ref{table:optimal_hyperparameters}.

Finally, we also conducted initial experiments on annealing schedule parameters (including reverse annealing) but abolished that route due to insignificant effects on the result.

\end{document}